\documentclass[twocolumn,groupedaddress,showpacs,showkeys]{revtex4}

\usepackage{graphicx}
\usepackage{amsmath}
\usepackage{amssymb}
\usepackage{amsfonts}
\usepackage{epsfig}

\def\be{\begin{equation}} \def\ee{\end{equation}}
\def\bea{\begin{eqnarray}} \def\eea{\end{eqnarray}}

\newcommand{\ket}[1]{| #1 \rangle}
\newcommand{\bra}[1]{\langle #1 |}

\begin{document}
\title{Coulomb matrix elements of bilayers of confined charge
  carriers with arbitrary spatial separation} 

\author{Ian Mondragon-Shem$^{\pi}$, Francisco E. L\'opez,$^{\dagger}$
  and Boris A. Rodr\'iguez$^{\pi}$} \email[Email:
]{banghelo@fisica.udea.edu.co} \affiliation{$\pi$ Instituto de
  F\'isica, Universidad de Antioquia, AA 1226 Medell\'in,
  Colombia\\ $^{\dagger}$ Centro de Investigaciones, Instituto Tecnol\'ogico de
  Medell\'{i}n }

\begin{abstract}
We describe a practical procedure to calculate the Coulomb matrix
elements of 2D spatially separated and confined charge carriers, which
are needed for detailed theoretical descriptions of important
condensed matter finite systems. We derive an analytical expression,
for arbitrary separations, in terms of a single infinite series and
apply a u-type Levin transform in order to accelerate the resulting
infinite series. This procedure has proven to be efficient and
accurate. Direct consequences concerning the functional dependence of
the matrix elements on the separation distance, transition amplitudes
and the diagonalization of a single electron-hole pair in vertically
stacked parabolic quantum dots are presented.
\end{abstract}

\maketitle

The interesting physical properties displayed by charge carriers in
spatially confining settings promise to have a diverse range of
applications relevant to technological developments in the near
future. Systems such as semiconductor quantum wells, quantum wires and
quantum dots, will have great impact in the progress of quantum
computation and quantum information \cite{Nielsen} and in producing
new technological devices in electronics, spintronics\cite{Zutic} and
optoelectronics \cite{Yariv}. Furthermore, on the experimental and
theoretical side, systems which have acquired increasing relevance
over the past decade, such as indirect excitons \cite{ButovBEC},
graphene bilayers \cite{Neto} and indirect magnetobiexcitons
\cite{Oleg}, show great promise for present-day and future
understanding of condensed matter systems.

In order to grasp the full potential of these advances, it is
necessary to have an increasingly detailed theoretical understanding
of the physics of confined charge carriers. These studies have
included, for example, learning about their optical properties,
describing their response to externally applied electric fields, and
understanding their behaviour when they are subject to external
magnetic fields \cite{Lerner,fqhe}.

Such theoretical studies have considered, in various levels of
approximation, the correlations introduced by the Coulomb interaction
among charge carriers. It is well known that the physics of confined
charge carriers is fundamentally affected by the Coulomb interaction
between charge carriers. As a consequence, a complete theoretical
study of the physical properties of charge carriers in confining
settings must take into account the full and non-trivial correlations
that arise from the long-ranged Coulomb interaction.

With this in sight, we present in this contribution a procedure to
efficiently compute the matrix elements of the Coulomb interaction
between charge carriers confined by a two dimensional parabolic
potential and which are spatially separated by a general interplane
distance.  This particular type of physical system is relevant in the
study of graphene bilayers\cite{Neto}, electron bilayers
\cite{Badalyan}, indirect biexcitons \cite{Meyertholen}, bilayer
quantum Hall systems \cite{bqhe}, self assembled quantum dots
\cite{Kuther}, coupled quantum wells \cite{Butov1999,Lozovik2005},
quantum dots formed by lateral fluctuations in the well
\cite{Butov1994}, among many others. Having an efficient way to
compute these elements opens the way for very relevant finite system
calculations (i.e. $O(10)-O(10^2)$ particles), as opposed to the types
of calculations that invoke the thermodynmic limit and that are
usually encountered in some theoretical formalisms. Among the
important finite system calculations, we may mention the Hartree-Fock,
the Hartree-Fock-Bogoliubov (e.g. through the use of BCS-type wave
functions) and the Random Phase Approximation approximations
\cite{Blaizot,Fetter}.
\begin{center}
\begin{figure}
\centering
\includegraphics[scale=.4]{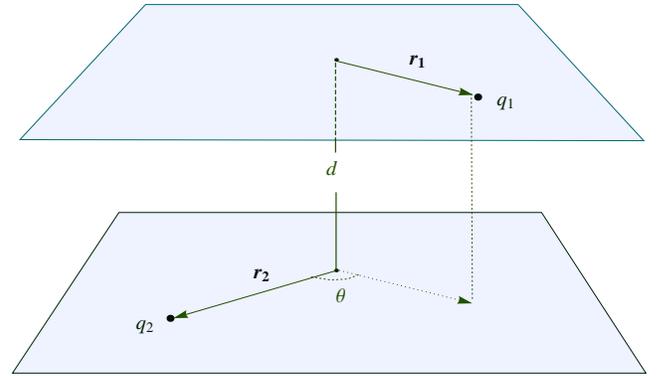}
\caption{Bilayer of interplane distance $d$, with charge carriers at
  $\mathbf{r_1}$ and $\mathbf{r_2}$, and relative angle
  $\theta$.}\label{systemfig}
\end{figure}
\end{center}
In the case of the matrix elements of confined charge carriers moving
in the same plane, the problem can be treated in an analytic form,
yielding a general expression for the Coulomb matrix elements
\cite{Halonen,Jacak}. However, when this formula is used for elements
that involve states with high quantum numbers of either angular
momenta or of radial excitations, convergence problems arise; it is
then necessary to use appropriate numerical methods to obtain reliable
results \cite{Boris,Ricardo,Alain,Augusto}. We shall see in this paper
that the computation of the matrix elements for the spatially
separated case also involves the use of a special numerical procedure
in order to accelerate the infinite series that arises in the
expansion of the expression for the Coulomb matrix elements.

This paper is organized as follows: in Sec. \ref{sec2}, we review the
Coulomb elements when the spatial separation $d$ is equal to zero. In
Sec. \ref{sec3} we give the expression for the Coulomb matrix elements
calculated for the $d\ne0$ case, followed by Sec. \ref{sec4} where the
method used to accelerate convergence is explained. Finally, in
Sec. \ref{sec5} we present some main results, and some conclusions are
given in Sec. \ref{sec6}. Appendix \ref{sec7} explains the notation
and basis ordering used throughout the article.

\section{Coulomb matrix element with $d=0$}\label{sec2}
The eigenfunctions $ \vert i \rangle = \vert n_{i},l_{i} \rangle $ of
a particle subject to either a two-dimensional harmonic potential \cite{Cohen} or a
perpendicular magnetic field \cite{Landau} are given by 
\begin{eqnarray}\label{wf}
\phi_{i}\left( \textbf{r} \right) = \phi_{n_{i},l_{i}}(r,\theta) =
C_{n_{i},\vert l_{i} \vert}r^{\vert l_{i} \vert}L_{n_{i}}^{\vert l_{i}
  \vert}\left( r^{2} \right)e^{-r^{2}/2}e^{\imath \theta l_{i}},
\end{eqnarray}
\noindent
with $n_{i}\in\{0\}\cup\mathbb{N}$, $l_{i}\in\mathbb{Z}$,
$L_{n_{i}}^{\vert l_{i} \vert}(r)$ are the usual associated Laguerre
polynomials and the normalization constant is given by $
C_{n_{i},\vert l_{i} \vert} = \sqrt{n_{i}! / \left[ \pi \left( n_{i}+
    \vert l_{i} \vert \right)! \right] } $; all lengths are scaled by
the relevant length scale of the system i.e. the harmonic oscillator
length or magnetic length, a choice which depends on the external
potential being considered. The corresponding energies are

\begin{equation}\label{enosc}
\epsilon_{i}^{osc}=\hbar \omega_0 (2n_i+\vert l_i \vert+1),
\end{equation}
for the harmonic oscillator basis and 
\begin{equation}\label{enland}
\epsilon_{i}^{landau}=\hbar \omega_0 (2n_i+\vert l_i\vert \pm l_i+1),
\end{equation}
for the Landau basis.  

Let $\bra{ij}V(d)\ket{kl}\equiv V(i,j,k,l,d)$ denote the Coulomb
interaction matrix element for spatially separated charge carriers,
with $d$ the length of the separation (see Fig. \ref{systemfig}). Then,
the matrix element between any two pairs of single-particle states for
the $d=0$ case is written as
\begin{eqnarray}\label{forma elemento directo}
&&\bra{ij}V(0)\ket{kl} = \langle ij \vert \frac{1}{\vert
  \textbf{r}_{1}-\textbf{r}_{2} \vert} \vert kl \rangle\nonumber \\ &
=&\int
\dfrac{d^{2}r_{1}d^{2}r_{2}}{\vert \textbf{r}_{1}-\textbf{r}_{2}
  \vert}\phi_{i}^{*}\left( \textbf{r}_{1} \right)\phi_{j}^{*}\left(
\textbf{r}_{2} \right)\phi_{k}\left( \textbf{r}_{1}
\right)\phi_{l}\left( \mathbf{r}_{2} \right).
\label{24}
\end{eqnarray}

\noindent
It should be noted that $\mathbf{r}_{i}$ is a 2D in-plane vector.

Many authors have found various expressions for these elements
\cite{Halonen,Jacak,Tsiper}. For example, using the
two-dimensional Fourier transform for the Coulomb potential
\begin{equation}
\frac{1}{|\textbf{r}_{1}-\textbf{r}_{2}|} = \int \frac{d^{2}q}{(2\pi q)}
e^{i\textbf{q}\cdot(\textbf{r}_{1}-\textbf{r}_{2})},
\end{equation}
it is possible to express the Coulomb matrix element in terms of
finite sums in the form \cite{Halonen}
\begin{eqnarray}\label{exact}
&&V(i,j,k,l,d=0)=\delta_{l_{in},l_{out}}C_{n_{i},l_{i}}C_{n_{j},l_{j}}C_{n_{k},l_{k}}C_{n_{l},l_{l}}
  \nonumber\\ & &\times
  \sum_{m_{i}=0}^{n_{i}}\sum_{m_{j}=0}^{n_{j}}\sum_{m_{k}=0}^{n_{k}}\sum_{m_{l}=0}^{n_{l}}\alpha_{ik}!\alpha_{jl}!\nonumber\\ &
  &\times
  \beta_{i}\beta_{j}\beta_{k}\beta_{l}\sum_{p=0}^{\alpha_{ik}}\sum_{s=0}^{\alpha_{jl}}\dfrac{(-1)^{p+s}(\alpha_{ik}+l_{ik})!}{(\alpha_{ik}-p)!(l_{ik}+p)!}\nonumber\\ &
  &\times
  \dfrac{(\alpha_{jl}+l_{jl})!}{(\alpha_{jl}-s)!(l_{jl}+s)!}\dfrac{\Gamma(l_{ij}+p+s+\frac{1}{2})}{p!s!2^{l_{ij}+p+s+\frac{1}{2}}}.
\end{eqnarray}

\noindent
The $\delta_{l_{in},l_{out}}$ expresses the conservation of the
angular momentum $l_{out}=l_{i}+l_{j}=l_{k}+l_{l}=l_{in}$. We have,
for convenience, defined the quantities $l_{ik}=\vert
l_{i}-l_{k}\vert=l_{jl}$, $\alpha_{ik}=(m_{i}+m_{k}+(\vert
l_{i}\vert+\vert l_{k}\vert-l_{ik})/2)$ and
$\beta_{i}=((-1)^{m_{i}}/m_{i}!)(\vert l_{i}\vert + n_{i})!/((\vert
l_{i}\vert + m)!(n_{i}-m_{i})!)$.  

Although expression (\ref{exact}) involves finite sums, it is
inefficient for computing matrix elements that involve states of high
angular momentum. The main difficulty arises from the computation of
the large factorials. This renders the calculation of the matrix
elements computationally expensive.

An alternative way of calculating the Coulomb matrix element is to
write
\begin{equation}
\frac{1}{|\textbf{r}_{1}-\textbf{r}_{2}|}=\frac{1}{\sqrt{\pi}}\int_{0}^{\infty}\frac{dt}{t^{1/2}}e^{-t|\textbf{r}_{1}-\textbf{r}_{2}|},
\end{equation}
By expanding the exponential inside the integral, replacing the result
in (\ref{forma elemento directo}) and integrating the angle and radial
variables,we arrive at an explicit infinite series. The remaining step
of calculating the infinite series was tackled successfully in
\cite{BorisPhd} by making use of a series acceleration algorithm
(different from the one used in this work).

\section{The bilayer case: Coulomb matrix element with $d\ne0$}\label{sec3}
Let us now proceed to derive an expression for the general case $d \ne
0$. The Coulomb interaction in this case reads
\begin{eqnarray}
  \dfrac{1}{\vert
    \textbf{r}_{1}-\textbf{r}_{2}+\textbf{d}\vert}&=&\dfrac{1}{\sqrt{(
    \textbf {r}_{1}-\textbf {r}_{2})^{2}+\textbf
            {d}^{2}}}\nonumber\\ &=&\dfrac{1}{\sqrt{r_{1}^{2}+r_{2}^{2}-2r_{1}r_{2}cos\theta+d^{2}}}\nonumber\\ &=&\dfrac{1}{\sqrt{\pi}}\int_{0}^{\infty}\dfrac{dt}{t^{1/2}}e^{-t(r_{1}^{2}+r_{2}^{2}+d^{2})}
\nonumber\\ &
&\times\sum_{n=0}^{\infty}\dfrac{2^{n}t^{n}r_{1}^{n}r_{2}^{n}cos^{n}(\theta)}{n!},
\end{eqnarray}

\noindent
$\theta$ being the angle between the 2D in-plane vectors
$\mathbf{r}_1$ and $\mathbf{r}_2$ (Fig. \ref{systemfig}), and the vector
$\mathbf{d}$ being defined as $\textbf{d}=d \widehat{k}$. Note that
$\mathbf{r}_1$, $\mathbf{r}_2$, and $\textbf {r}_{1}-\textbf {r}_{2}$
are perpendicular to $\textbf {d}$.  Manipulations analogous to those
in the previous case, and use of $L_{n}^{\vert l
  \vert}(r)=\sum_{m=0}^{n}((-1)^{m}/m!)(\vert l\vert + n)!
r^{m}/((\vert l\vert + m)!(n-m)!)$ lead to the following expression
for the Coulomb matrix elements
\begin{eqnarray}\label{indirecto}
\bra{i,j}&V(d)&\ket{k,l}=\delta_{l_{in},l_{out}}C_{n_{i},l_{i}}C_{n_{j},l_{j}}C_{n_{k},l_{k}}C_{n_{l},l_{l}}\nonumber\\ &
\times&
\sum_{m_{i}=0}^{n_{i}}\sum_{m_{j}=0}^{n_{j}}\sum_{m_{k}=0}^{n_{k}}\sum_{m_{l}=0}^{n_{l}}
\beta_{i}\beta_{j}\beta_{k}\beta_{l}\nonumber\\ &\times&
\sum_{p=0}^{\infty}\dfrac{(\alpha_{ik}+p)!(\alpha_{jl}+p)!}{p!(p+l_{ik})!}
\nonumber\\ &\times&\left\{
_{1}F_{1}(1/2+2p+l_{ik},-\eta-l-1/2,d^{2})\right. \nonumber\\ &\times&
\left. \dfrac{\Gamma(1/2+2p+l_{ik})\Gamma(\eta+l+3/2)}{\Gamma(2+\eta+2p+l_{ik}+l)}\right.\nonumber\\&+&\left. d^{3+2\eta+2l}\Gamma(-\eta-l-3/2)
\right.\\ &\times&
\left. _{1}F_{1}(2+\eta+2p+l_{ik}+l,\eta+l+5/2,d^{2})\right\} \nonumber,
\end{eqnarray}

\noindent
where $2l=\vert l_{i} \vert+\vert l_{j} \vert+\vert l_{k} \vert+\vert
l_{l} \vert$, $\eta=m_{i}+m_{j}+m_{k}+m_{l}$ and $_{1}F_{1}(a,b,z)$ is
the confluent hypergeometric funtion \cite{Abramowitz}. This
expression reduces in an equivalent form to (\ref{exact}) when $d \to
0$ \cite{BorisPhd}.  Although our expression involves an infinite sum,
we can apply a numerical convergence procedure in order to make the
computation of the matrix elements more tractable.  To the best of our
knowledge, there is no expression nor procedure for a practical
calculation of the Coulomb matrix elements of spatially separated
charge carriers in a two dimensional confining potential.  As
mentioned in the introduction, this result is of significant practical
importance in finite system studies in various condensed matter
systems.

Incidentally, we would like to note first that it is possible to
reduce the matrix elements satisfying $l_i-l_j=l_k-l_l$ to only radial
integrals of complete elliptic functions of the first kind $K(k)$, as
long as $d>0$. Such a simplification is not possible when $d=0$ mainly
because of the singular behaviour of the Coulomb interaction in such a
case. The resulting integrals read
\begin{eqnarray}\label{consistency}
&& \bra{ij}V(d)\ket{kl}=\nonumber\\&&\delta_{l_{in},l_{out}}
  \int^{\infty}_0dr_1\int^{\infty}_0dr_2G(i,j,k,l,r_1,r_2;d), \end{eqnarray}
having defined the radial function
\begin{eqnarray}
&G&(i,j,k,l,r_1,r_2;d)=2\pi
  C_{n_il_i}C_{n_jl_j}C_{n_kl_k}C_{n_ll_l}\nonumber\\&&\times
  r_1^{\vert l_i\vert+\vert l_k\vert+1}r_2^{\vert l_j\vert+\vert
    l_l\vert+1}e^{-(r_1^2+r_2^2)}\nonumber\\&&\times L_{n_i}^{\vert
    l_i\vert}(r_1^2)L_{n_j}^{\vert l_j\vert}(r_2^2)L_{n_k}^{\vert
    l_k\vert}(r_1^2)L_{n_l}^{\vert
    l_l\vert}(r_2^2)\\&&\times\left(\frac{2 K\left(-\frac{4
      r_1r_2}{d^2+(r_1-r_2)^2}\right)}{\sqrt{(d^2+(r_1-r_2)^2)}}+\frac{2
    K\left(-\frac{4
      r_1r_2}{d^2+(r_1+r_2)^2}\right)}{\sqrt{(d^2+(r_1+r_2)^2)}}\right).\nonumber
\end{eqnarray}
Although the radial integrals have convergence problems for high
values of the radial quantum numbers, due to the oscillatory nature of
the Laguerre polynomials, the elements that can be computed agree with
our results which make use of the series acceleration algorithm to be
discussed presently.

\section{The series acceleration} \label{sec4}

We calculate the infinite series in the $p$ index that appears in
(\ref{indirecto}) using a u-type Levin transform to accelerate the
convergence \cite{Levin1}.  The basic idea is to construct an
alternate series which converges much faster than the one we have at
hand.  Intuitively, this could be done if there was a way to simulate
the asymptotic behaviour of the remainder $R_m$ of the series for
large values of $m$. We can achieve this by constructing functions
$\omega_m$ such that
\begin{equation}\label{asymp}
\lim_{m\rightarrow\infty}\frac{R_m}{\omega_m}=c,
\end{equation}
$c$ being of order unity. Having done this, we can then write
\begin{equation}
R_m=\omega_m \mu_m,
\end{equation}
such that the coefficients $\mu_m$ satisfy $\lim_{m\rightarrow
  \infty}\mu_m= c$. By doing this, the asymptotic behaviour of the
series will have been coded into the $\omega_m$ quantities.  Now we
must find a prescription for the coefficients $\mu_m$. This can be
done by writing, for large $m$, the expression
\begin{equation}
  \mu_m\approx\sum^{\infty}_{i=0}C_{i}\psi_{i}(m).
\end{equation}
This expression is reasonable so long as the functions $\psi_i(m)$
satisfy three conditions: first, we must have $\psi_0(m)=1$ and
$C_0=c$ $\forall m$, so that (\ref{asymp}) holds; second, for $i>0$ we
must have $\psi_i(m) \to 0$ when $m \to \infty$; and third we must
require that $\psi_{i+1}(m)=\mathcal{O}(\psi_{i}(m))$ , so that, in
taking the large $m$ limit, all terms $i\ne0$ vanish at the same
rate. The constants $C_{i}$ are still unknown at this stage of the
derivation.

With these definitions, the $m^{th}$ partial sum $s_m$ of our series
can be written as, for large $m$,
\begin{equation}
  s_m\approx s+\omega_m\sum^{\infty}_{i=0}C_{i}\psi_{i}(m).
\end{equation}
Here, $s$ is the exact value of the series we want to
compute. Finally, we truncate the sum, thus eliminating asymptotic
terms:
\begin{equation}\label{series_result}
\sigma_m=\sigma+\omega_m\sum^{k-1}_{i=0}C_{i}\psi_{i}(m).
\end{equation}
The corresponding change of notation from the exact $s$ to the
truncated $\sigma$ should be clear. Notice that the expression
(\ref{series_result}) tells us that the value $\sigma$ is such that
all $k+1$ equations, for $m$ running from $m$ to $m+k$, hold
simultaneously. That is to say, the truncated value $\sigma$ depends
on all terms which add up to $\sigma$ itself and, thus, it is as if it
``feelt'' the overall behaviour of the most important terms of the
series. In this sense, the value calculated through the use of the
u-type Levin transform can be thought of as an extrapolation of the
final sum that uses information from only the first few terms of the
complete series. The more terms we keep in the truncation, the better
the extrapolation.

In order to find the approximation $\sigma$ to the full sum, it is
necessary to solve the $k+1$ equations for the $k+1$ unknowns $\sigma$
and $\{C_i\}$. Hence, the resulting problem of calculating the series
has been reduced to computing the solution of a $k+1$ dimensional
matrix inversion problem. The larger the value of $k$, the better the
approximation to the exact value. The u-type Levin transform uses the
choices $\psi_{i}(m)=(m+\beta)^{-j}$ \cite{Levin1} and $\omega_n=n
a_n$, which have proven to behave well for a large family of series
\cite{Levin2}. The parameter $\beta$ is a real number that can be
chosen to improve the rate of convergence. 

As an example, in our calculations, an element such as
\begin{equation}
\bra{15}V(d=0.2)\ket{23}\approx -0.0747592223
\end{equation}
required $1635$ terms of the series to achieve convergence up to the
tenth decimal place using the term-by-term sum, whereas it required
less than $100$ terms using the Levin transform, and was calculated in
half the time using a standard linear solver. It should be noted that
the number of terms needed to achieve convergence will depend on the
matrix element that is being calculated and that the calculation time
will depend on the machine and the procedure used to solve the linear
system that results from the Levin transform.

\begin{center}
\begin{figure}
\includegraphics[scale=0.35]{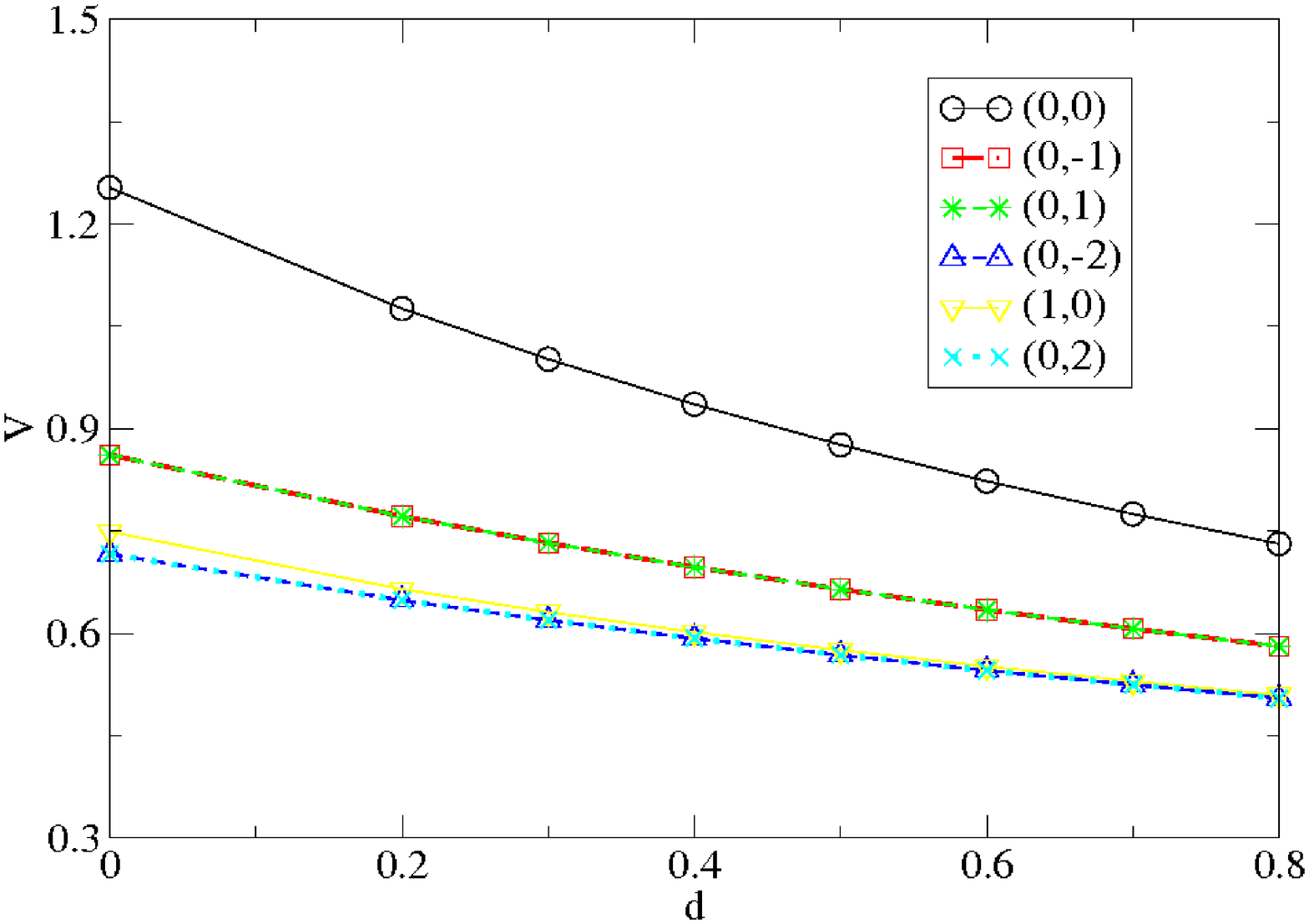}
\includegraphics[scale=0.35]{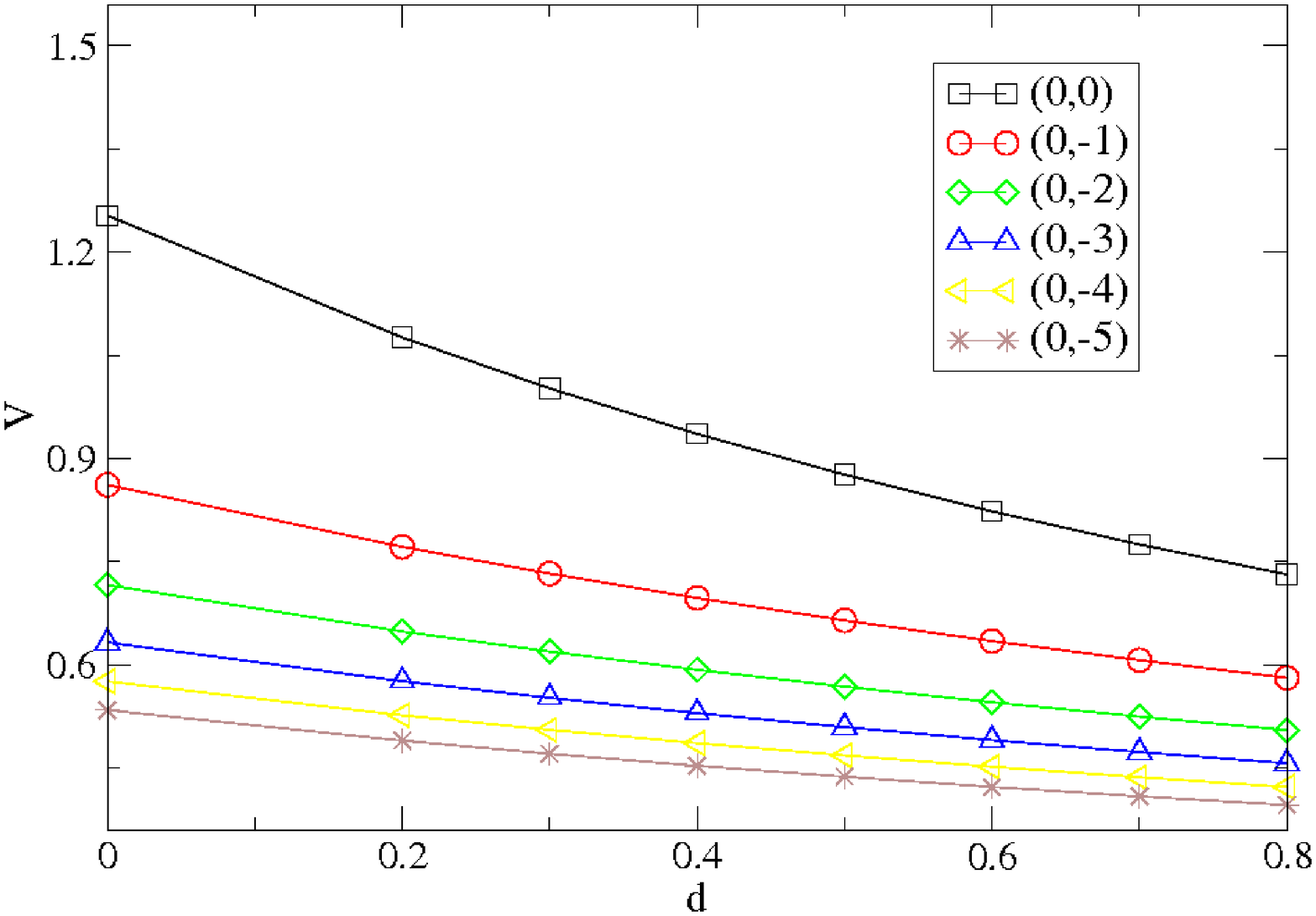}
\caption{ Elements of the form $\bra{i,i}V(d)\ket{i,i}$ for:
  \textbf{(Top)} In the ordering of the harmonic oscillator
  basis. \textbf{(Bottom)} In the ordering of the Landau basis. The
  parentheses denote the quantum numbers $(n,l).$}\label{fig_diag}
\end{figure}
\end{center}

\section{Main results}\label{sec5}

\subsection{\textbf{Comparison between the $d = 0$ and $d \ne 0$ cases}}

We first evaluate some particular matrix elements in order to get a
feeling of how they vary as functions of the distance $d$.

In Fig. (\ref{fig_diag}), we show the behavior of the Coulomb matrix
elements of the form $\bra{i,i}V(d)\ket{i,i}$ as functions of the
separation between the planes, for both the oscillator and Landau
basis. For both basis sets, there is a general decaying
behaviour. Clearly, when $d\gg 1$, all elements must approach
zero. Refer to the appendix for the notation used for the basis sets.

Although, there is a one-to-one correspondence between both basis
sets, there is a physical reason which justifies showing the
calculations in both basis. The Landau basis corresponds to the
solution of a charged particle moving in an external magnetic field
perpendicular to the plane of motion. This system is infinitely
degenerate with respect to the angular momentum quantum number and,
hence, its ordering is conceptually different to the finite number of
degenerate energy levels of the two dimensional harmonic oscillator
system. Thus, showing the calculation in each basis set is relevant
for each of the two different types of in-plane confinement.

\begin{center}
\begin{figure}
\includegraphics[scale=0.33]{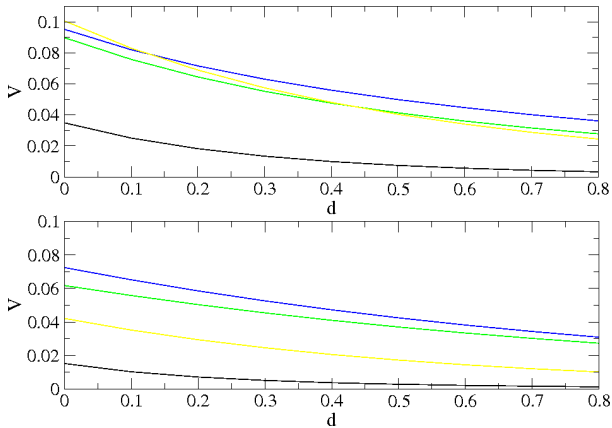}
\caption{Behaviour of the Coulomb average of randomly chosen pairs of
  states of total angular momentum $l=0$ \textbf{(top)} and $l=1$
  \textbf{(bottom)} as a function of $d$. This was done with the
  oscillator ordering.}\label{fig_angmom}
\end{figure}
\end{center}

Next, we construct a set of elements which involves pairs of states
with fixed total angular momentum (i.e. using sets of two-particle
states with some predetermined value of their total angular
momentum). In order to do this, such states were randomly chosen for
the cases $l=0$ and $l=1$. This is shown in
Fig. (\ref{fig_angmom}). The general exponentially decaying behaviour
observed in these plots seems to suggest that there is a scaling
behaviour behind elements with predetermined total angular
momentum. 

\begin{center}
\begin{figure}
\includegraphics[scale=.32]{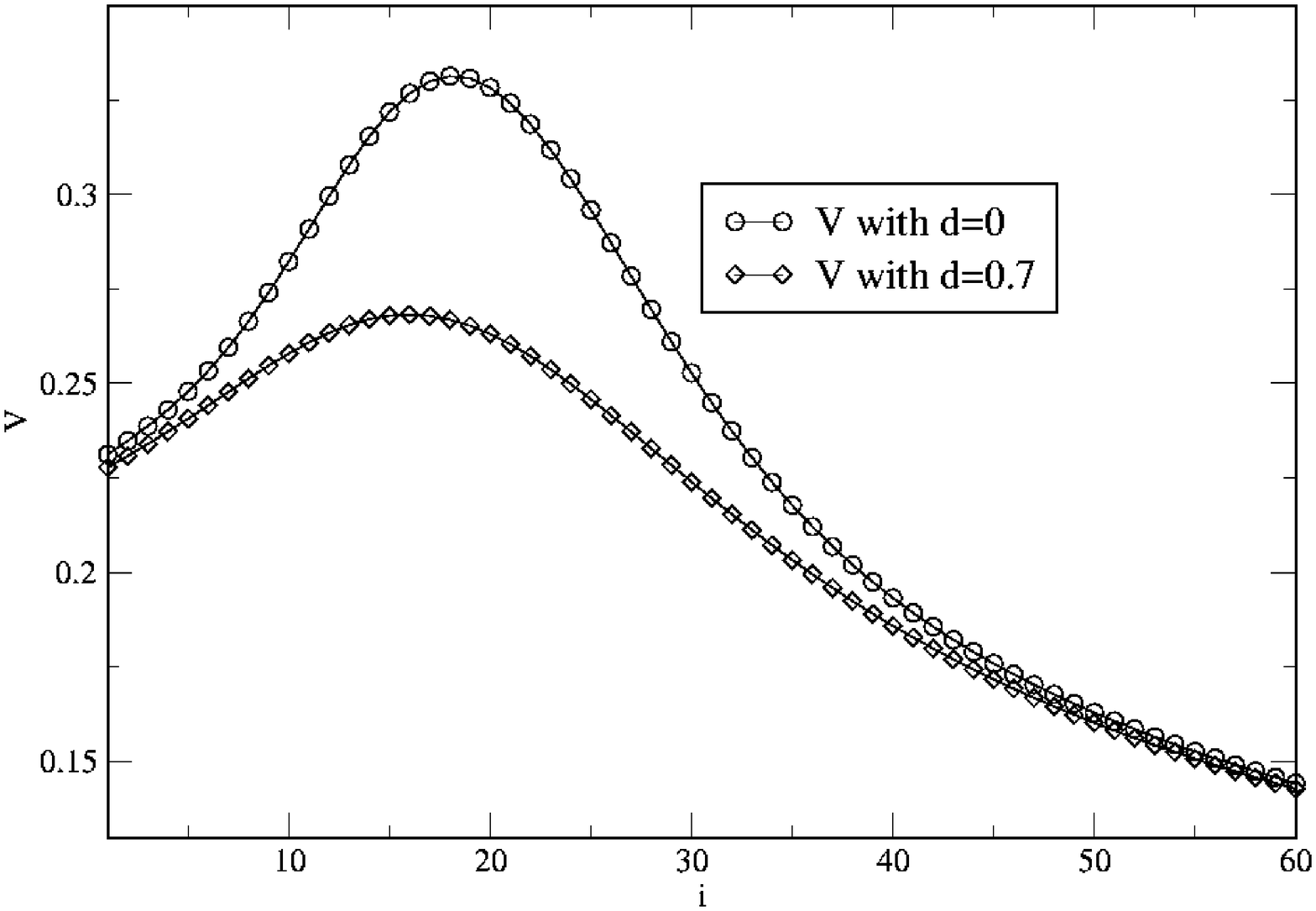}
\includegraphics[scale=.32]{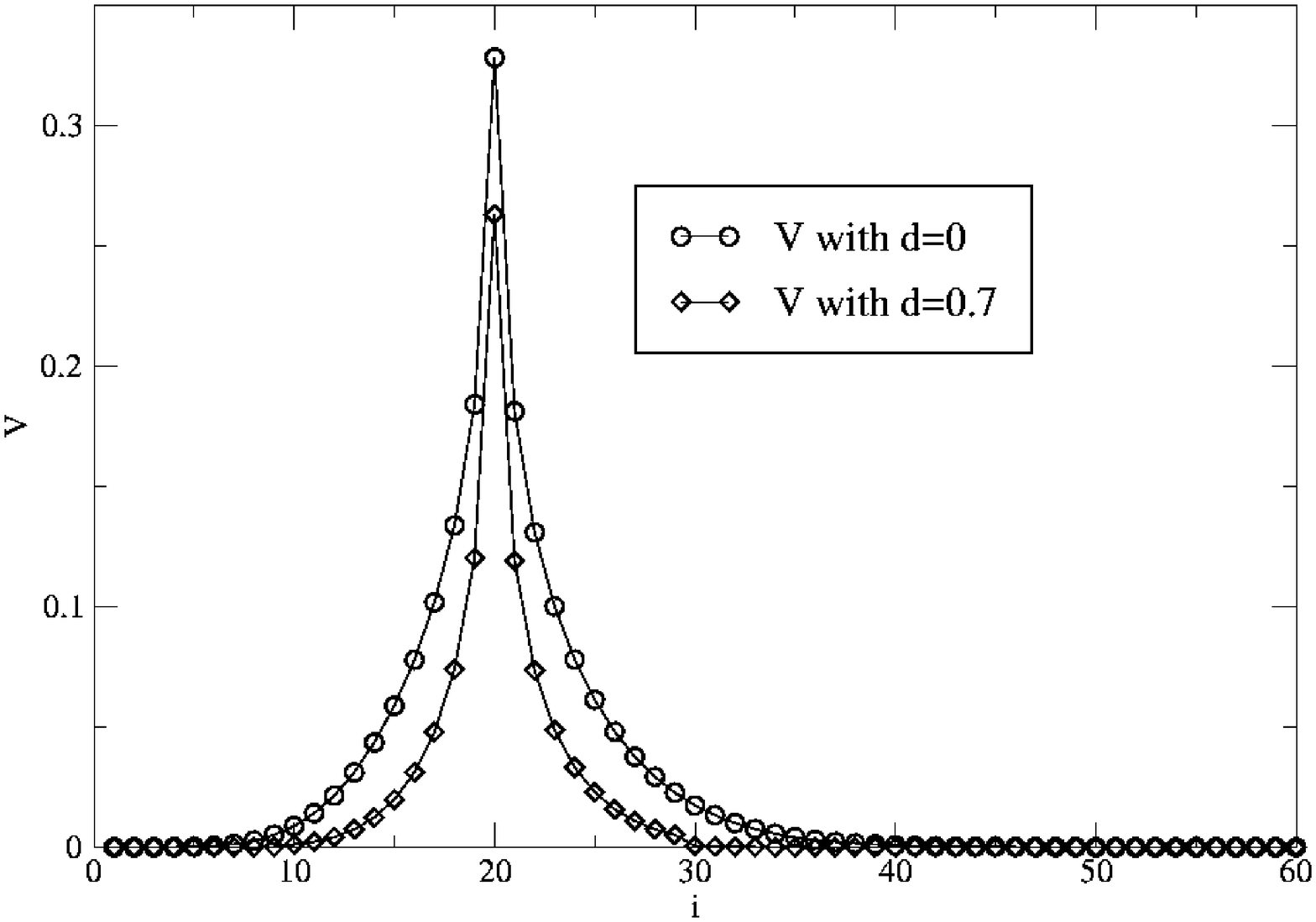}
\caption{\textbf{Top:} Direct Coulomb matrix element of the form
  $V(i,20,i,20,d)$. \textbf{Bottom:} Exchange Coulomb matrix element
  of the form $V(i,20,20,i,d)$. Both plots were done with the index
  $i$ running through the first 60 states of the first Landau
  level}\label{directindirect}
\end{figure}
\end{center}

This last observation leads to an interesting behaviour which is worth
emphasizing. Although all elements get suppressed in general as the
distance is increased, they do not all approach zero at the same
rate. Such a behaviour can be seen clearly in the top plot of
Fig.(\ref{fig_angmom}): one of the curves crosses two other curves,
indicating that it is decaying faster than the other two. This tells
us that, as $d$ varies, the entries in the matrix representation of
the Coulomb interaction vary in nontrivial ways with respect to each
other.  In the very large $d$ limit i.e. when $d\gg 1$, we may
obtain an expansion for the rate of decay
\begin{eqnarray}
&&\frac{\partial}{\partial d}V(i,j,k,l,d)\approx -\frac{1}{d^2}
  \delta_{ik}\delta_{jl},
\end{eqnarray}
so that not all elements tend asymptotically to zero at the same
rates, the rate depending on the indices of the states used in the
matrix element. Hence it is possible, by changing the distance between
the planes, for a system of charge carriers to exhibit different types
of dynamical behaviour, as opposed to just interacting in a kind of
screened Coulomb potential produced by the separation.
In Fig. (\ref{directindirect}) we show the most relevant elements for
the dynamics of Coulombian systems, namely $\bra{ij}V(d)\ket{ij}$ and
$\bra{ij}V(d)\ket{ji}$, the so-called direct and exchange terms, using
the first energy shell of the Landau basis. Note how, as we increase
the interplane distance, the direct terms get more suppressed,
relative to the $d=0$ value, than the exchange ones. Also, the
exchange elements decay faster than the direct ones as we change the
index $i$, a behaviour that is general for every $d$. This decay is
expected because the exchange elements involve integrals of amplitudes
which can interfere destructively when summed over, as opposed to the
direct elements which involve probability densities which are always
positive. Finally, the maximum value of the direct term is actually
shifted from its diagonal value $V(i,i,i,i,d)$, and this maximum gets
further shifted as $d$ increases. However, the maximum for the
exchange plot remains at the diagonal point, although this is actually
again the same diagonal (and, thus, direct) matrix element $V(i,i,i,i,d)$.

\subsection{\textbf{Amplitude transitions and diagonaliation of a single indirect exciton}}

\begin{center}
\begin{figure}
\includegraphics[scale=0.34]{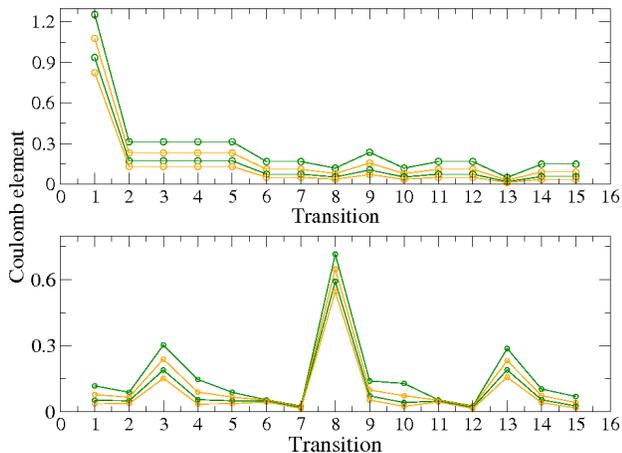}
\caption{Absolute value of the transition amplitudes between harmonic
  oscillator states of zero total angular momentum. Each continuous
  line has a fixed $d$ value.  As $d$ increases, the lines get
  suppressed. The initial state corresponds to two particles in
  $\ket{1}\ket{1}$ (\textbf{top}) and $\ket{4}\ket{6}$
  (\textbf{bottom}). The final two-particle states are ordered
  horizontally according to the list: $\ket{1}\ket{1}$,
  $\ket{1}\ket{5}$, $\ket{2}\ket{3}$, $\ket{3}\ket{2}$,
  $\ket{5}\ket{1}$, $\ket{2}\ket{9}$, $\ket{3}\ket{8}$,
  $\ket{4}\ket{6}$, $\ket{5}\ket{5}$, $\ket{6}\ket{4}$,
  $\ket{8}\ket{3}$, $\ket{9}\ket{2}$, $\ket{7}\ket{10}$,
  $\ket{8}\ket{9}$, $\ket{9}\ket{8}$.}\label{fig_trans}
\end{figure}
\end{center}

Since the Coulomb matrix elements connect two pairs of two-particle
states through the Coulomb interaction, we can use them to weigh how
probable it is for a particular transition to occur as a function of
the distance $d$. In order to study a simple transition, we computed
the matrix elements for a given pair of initial states and several
pairs of final states, all pairs having zero total angular
momentum. This is shown in Fig. (\ref{fig_trans}). Each continuous
line is the absolute value of the transition amplitude for a given
distance $d$ and each point represents a final state, as explained in
the caption of the figure. The succession of lines go from $d=0$
through $d=1$. As $d$ increases, all points (i.e. all considered
transitions) generally get suppressed.
Once again, we note that the rate at which each coefficient gets
suppressed is not the same for all, the diagonal transition being
typically the most significant one, as further calculations of other
similar cases have shown. In fact, when $d$ is of order $1$ (i.e. when
the distance is of the order of the oscillator length), almost all of
the coefficients are negligible, except for the diagonal ones. This
reasonably shows that, when the planes are separated beyond the
characteristic distance of the system, all transitions get suppressed,
albeit at unequal rates.

\begin{center}
  \begin{figure}[h]
    \includegraphics[scale=0.33]{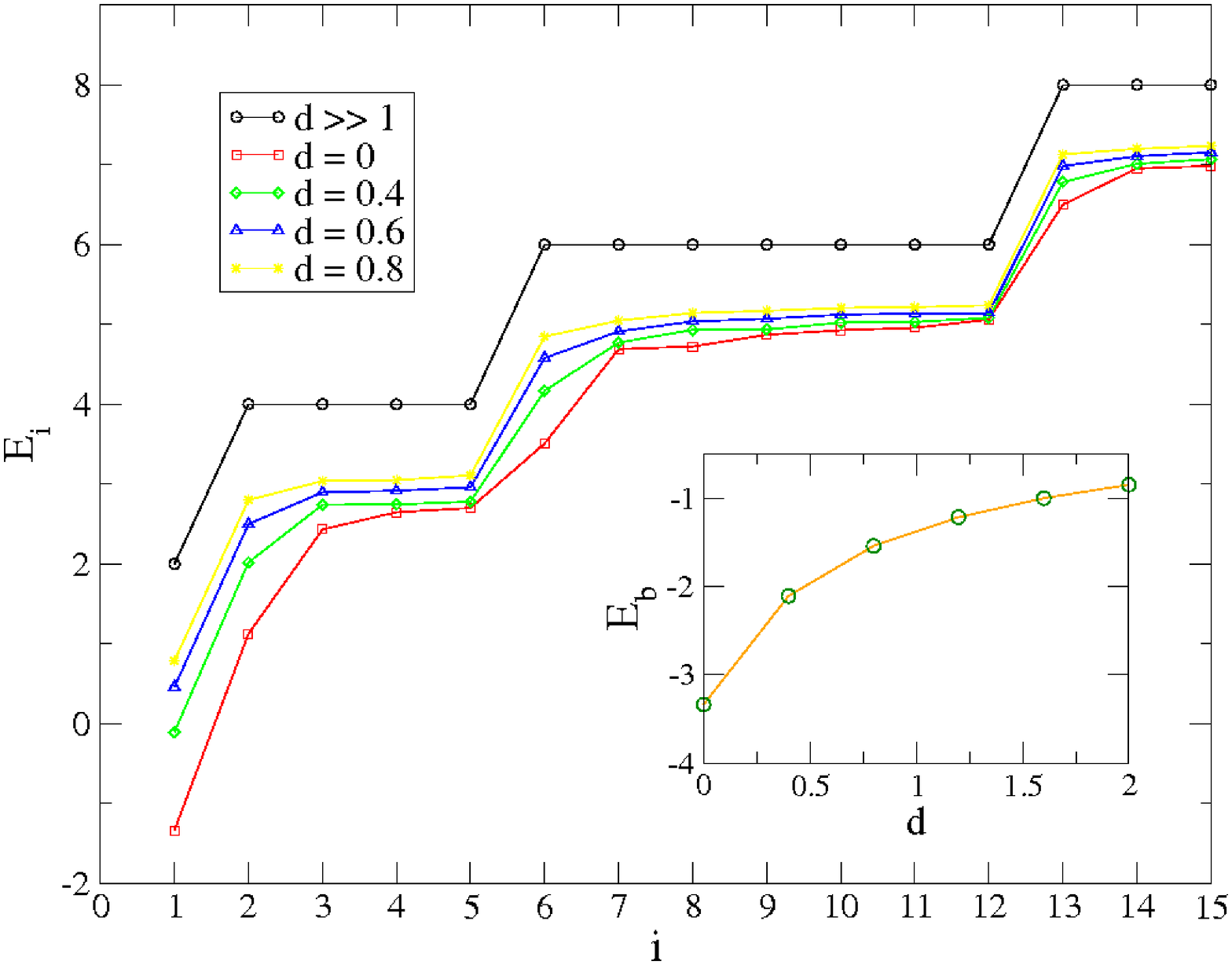}
    \caption{First few exciton energies $E_i$ for zero total angular
      momentum. Each continuous line represents a fixed value of
      $d$. The inset shows the binding energy, defined as
      $E_b=E_1-E_{osc}$ (i.e. the portion of the energy of the bound
      electron-hole pair which is due solely to the Coulomb
      interaction), as a function of the interplane distance
      $d$}\label{fig_couldiag}
  \end{figure}
\end{center}

Finally, as a more physical application, we have considered the simple
case of an electron and a hole spatially separated in an effective
mass hamiltonian with a harmonic potential. The eigenstates of this
system are those of a single spatially indirect exciton. The
hamiltonian, in dimensionless units, reads
\begin{equation}
H=\sum_i\{\epsilon^{(e)}_i e^{\dagger}_ie_i+\epsilon^{(h)}_ih^{\dagger}_ih_i\}-\beta
\sum_{i\overline{j}k\overline{l}}\bra{ij}\ket{kl}e^{\dagger}_ih^{\dagger}_{\overline{j}}h_{\overline{l}}e_k.
\end{equation}
In this expression, we have made the convenient definition
$\bra{ij}\ket{kl}=\bra{ij}V(d)\ket{kl}$. The $\beta$ parameter denotes the ratio between the Coulomb and harmonic oscillator energies, and the $\epsilon^{(e)}_i$ and $\epsilon^{(h)}_i$ are the dimensionless harmonic oscillator energies of the electron and the hole, respectively. The diagonalization for the $l=0$ block of the hamiltonian is shown in
Fig.(\ref{fig_couldiag}). As before we notice the unequal rates of
change as $d$ is varied only this time it is evidenced with the
eigenenergies of the exciton. As $d$ increases, the spectrum tends to the harmonic oscillator energies for two particles with total zero angular momentum, as expected.

\subsection{\textbf{Convergence sum rule}}
As a consistency check of the above results, we computed an identity
involving the Coulomb matrix elements, reminiscent of the sum rules encountered in atomic physics. Namely, it can be checked that,
provided $l_i-l_j=l_k-l_l$, we can integrate out the angular integrals
exactly to obtain a relation of the form
\begin{eqnarray}\label{consistency}
&&\bra{ij}V^2(d)\ket{ij}=\sum_{k l}\vert \bra{ij}V(d)\ket{kl} \vert^2=\\&&\delta_{l_{in},l_{out}}
\int^{\infty}_0dr_1\int^{\infty}_0dr_2F(i,j,k,l,r_1,r_2;d),\nonumber \end{eqnarray}
with
\begin{eqnarray}
&F&(i,j,k,l,r_1,r_2;d)=C_{n_il_i}C_{n_jl_j}C_{n_kl_k}C_{n_ll_l}\nonumber\\&&\times
  r_1^{\vert l_i\vert+\vert l_k\vert+1}r_2^{\vert l_j\vert+\vert
    l_l\vert+1}e^{-(r_1^2+r_2^2)}\nonumber\\&&\times L_{n_i}^{\vert
    l_i\vert}(r_1^2)L_{n_j}^{\vert l_j\vert}(r_2^2)L_{n_k}^{\vert
    l_k\vert}(r_1^2)L_{n_l}^{\vert
    l_l\vert}(r_2^2)\nonumber\\&&\times\frac{(2\pi)^2}{\sqrt{(d^2+(r_1-r_2)^2)(d^2+(r_1+r_2)^2)}}.
\end{eqnarray}

\begin{center}
  \begin{figure}[h]
    \includegraphics[scale=0.33]{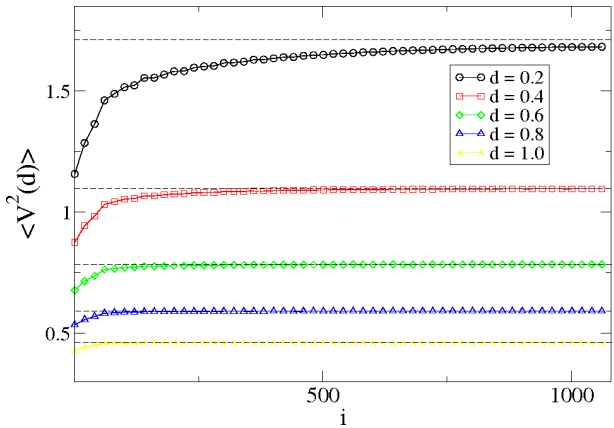}
    \caption{Convergence of the left hand side of
      Eq.(\ref{consistency}) as a function of the number of oscillator
      states $\ket{kl}$ used in the summation, starting from the two
      particle states which are closer to the state $\ket{ij}$ (the
      $i$ in horizontal axis is an index that denotes the $i^{th}$
      element of the two-particle oscillator basis of the form
      $\ket{kl}$). In this particular example, the state $\ket{ij}$ is
      the ground state. The horizontal dashed lines are the exact
      values calculated with the right hand side of
      Eq.(\ref{consistency}).}\label{fig_convergence}
  \end{figure}
\end{center}

Note that the convergence of the curve corresponding to $d=0.2$ is
slower than that of the other curves. This has a simple and clear
physical meaning. As the spatial separation between the planes becomes
smaller, it will be easier for the Coulomb interaction to induce
transitions starting from two particles in the oscillator ground state
to two particles in two other higher oscillator states. This means
that all amplitudes that we add in the sum rule become more
significant as $d$ becomes smaller and, thus, the sum will require
more and more terms which involve higher oscillator
states. Notwithstanding this, there is clear convergence.

The radial integrals can be evaluated numerically by various methods,
which makes the computation of the right hand side of
(\ref{consistency}) completely independent from our main numerical
approach, which is in turn used to compute the left hand side. We
found that the relation (\ref{consistency}) was very well satisfied,
as can be seen in Fig. (\ref{fig_convergence}).

\begin{center}
\begin{figure}
\includegraphics[scale=.32]{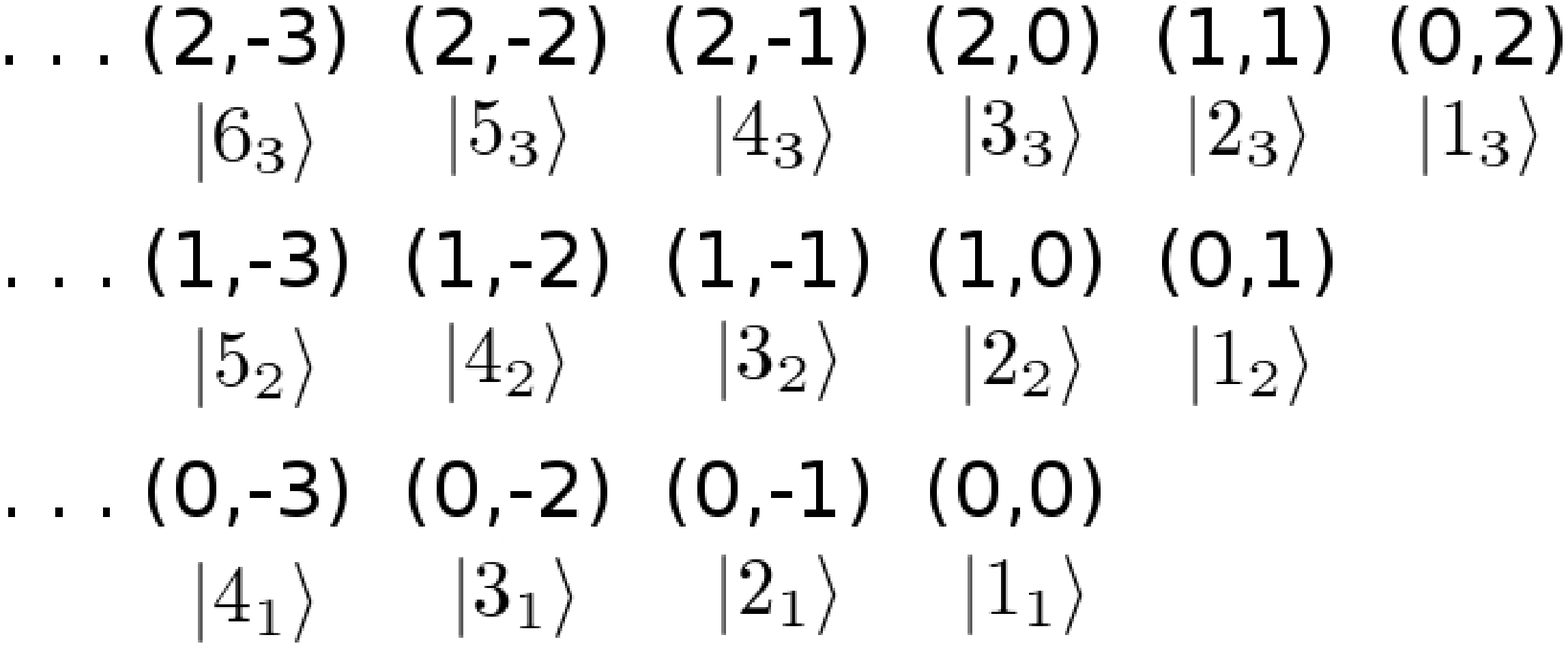}
\caption{Ordering of the Landau basis.  Each line is an energy shell
  which has infinite many elements. The notation $\ket{N_{ll}}$
refers to the $N^{th}$ state of the $ll$ Landau
level.}\label{basis_landau}
\includegraphics[scale=.27]{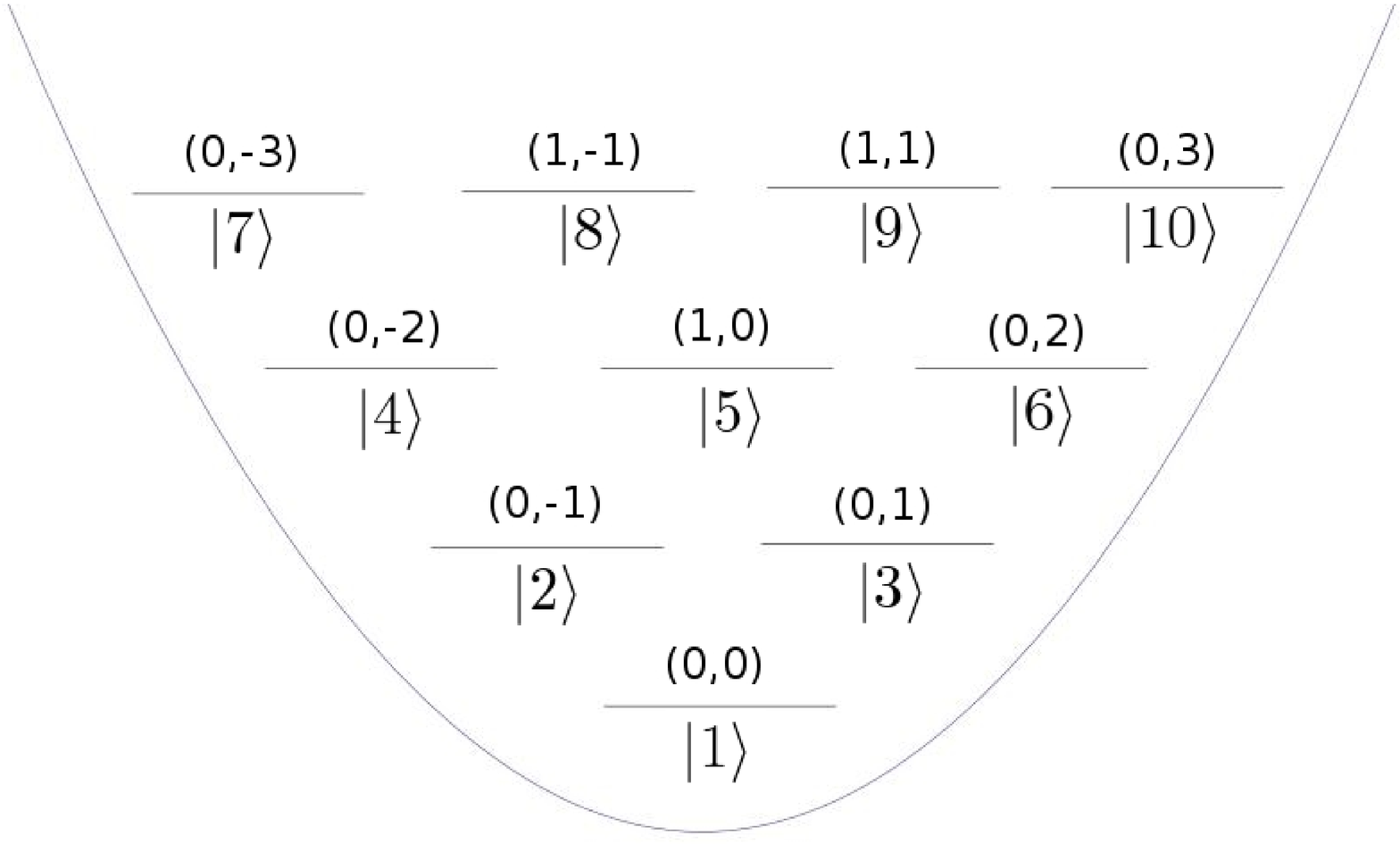}
\caption{Ordering of the harmonic oscillator basis. Each line is an
  energy shell which has finite many
  elements.}\label{basis_oscillator}
\end{figure}
\end{center}

\section{Conclusions}\label{sec6}

In the work presented here, we have shown the procedure and results of
a numerical calculation of the Coulomb matrix elements of spatially
separated charge carriers confined either by a two dimensional
parabolic quantum dot or through the use of a magnetic field. We have
found that, in a very important way, each matrix element is affected
differently as $d$ is varied. This could lead to different and
interesting physical regimes of a system subject to interparticle
Coulomb interaction.

Although the procedure we have implemented above has worked quite
well, it should be noted that, as with any numerical algorithm, care
must be taken in evaluating elements with too high a difference in
quantum numbers of the involved states e.g. high differences between
the $l_i$ and $l_k$ or $k_j$ and $k_l$. These cases can become
pathological because of the highly oscillatory nature of the
associated wavefunctions.

We would like to emphasize the relevance of the procedure presented
here. With such elements, theoretical studies can be carried out that
are comparable with experimental systems. Furthermore, this study is
relevant for describing and understanding many-body effects such as
collective behaviour and quantum correlations.

Work that involves the in-depth study of spatially indirect excitons
within a finite system framework using Hartre-Fock and BCS type
approximations is already on the way. For such a study, the matrix
elements computed in the present paper will be critically needed.\\

\begin{acknowledgements}
The authors would like to thank the CODI - Univ. de Antioquia and the
Centro de Investigaciones del ITM, for partial financial
support. B.A.R. acknowledges discussions with R. Perez, A. Delgado and
A. Gonzalez concerning the calculation of the $d=0$ case. The authors
are very grateful for enlightning discussions with J. Mahecha,
H. Vinck and C. Vera.
\end{acknowledgements}

\section{Appendix: basis ordering}\label{sec7}
Since we will be refering to two different basis sets, we will specify
the ordering used throughout the paper. Both basis sets are organized
in shells according to their respective energies given in
Eq.(\ref{enosc}) and Eq.(\ref{enland}). For the Landau basis, the
logic of the numbering can be extracted from Fig.(\ref{basis_landau});
and that of the oscillator basis, from
Fig.(\ref{basis_oscillator}). The parentheses in these figures follow
the notation $(n,l)$ i.e. the radial and the angular momentum quantum
numbers associated to the corresponding wavefunctions shown in
Eq.(\ref{wf}) of the charge carriers.

\end{document}